\documentclass{elsart}

\usepackage{graphicx}
\usepackage{amsmath}
\usepackage{amssymb}

\begin{document}

\begin{frontmatter}

\title{Order-$\mathbf N$ Cluster Monte Carlo Method for Spin Systems
  with Long-range Interactions}

\author[label1]{Kouki Fukui}
\ead{monta@looper.t.u-tokyo.ac.jp}
and
\author[label1,label2]{Synge Todo}
\ead{wistaria@ap.t.u-tokyo.ac.jp}
\address[label1]{Department of Applied Physics, University of Tokyo,
  Tokyo 113-8656, Japan}
\address[label2]{CREST, Japan Science and Technology Agency,
  Kawaguchi, 332-0012, Japan}

\begin{abstract}
  An efficient $O(N)$ cluster Monte Carlo method for Ising models with
  long-range interactions is presented.  Our novel algorithm does not
  introduce any cutoff for interaction range and thus it strictly
  fulfills the detailed balance.  The realized stochastic dynamics is
  equivalent to that of the conventional Swendsen-Wang algorithm,
  which requires $O(N^2)$ operations per Monte Carlo sweep if applied
  to long-range interacting models.  In addition, it is shown that the
  total energy and the specific heat can also be measured in $O(N)$
  time.  We demonstrate the efficiency of our algorithm over the
  conventional method and the $O(N \log N)$ algorithm by Luijten and
  Bl\"ote.  We also apply our algorithm to the classical and quantum
  Ising chains with inverse-square ferromagnetic interactions, and
  confirm in a high accuracy that a Kosterlitz-Thouless phase
  transition, associated with a universal jump in the magnetization,
  occurs in both cases.
\end{abstract}

\begin{keyword}
% keywords here, in the form: keyword \sep keyword
long-range interaction \sep cluster algorithm \sep $O(N)$ method \sep Ising model \sep quantum Monte Carlo \sep Kosterlitz-Thouless transition
% PACS codes here, in the form: \PACS code \sep code
\PACS 02.50.Ga \sep 02.70.Ss \sep 05.10.Ln \sep 64.60.De
\end{keyword}
\end{frontmatter}

%%%%%%%%%%%%%%%%%%%%%%%%%%%%%%%%%%%%%%%%%%%%%%%%%%%%%%%%%%%%%%%%%%%%%%%%%%%%%%%

\section{Introduction}

Systems with long-range interactions exhibit more involved phase
diagrams and richer critical phenomena than those with only
nearest-neighbor interactions.  One of the most prominent examples is
the Ising model with long-range interactions, whose Hamiltonian is
defined as
\begin{equation}
  - \beta  {\mathcal H} = \sum_{i<j} \beta J_{ij} \sigma_i^z \sigma_j^z , 
  \label{eqn:hamiltonian}
\end{equation}
where $\sigma_i^z = \pm 1$, $J_{ij}$ is the coupling constant between
the $i$th and $j$th sites ($i,j = 1,2, \cdots ,N$), and $N$ is the
total number of spins.  Among the models described by
Eq.~(\ref{eqn:hamiltonian}), the one-dimensional chain with
algebraically decaying interactions has been studied most intensely so
far.  The interaction $J_{ij}$ for the model is written as
\begin{equation}
  J_{ij} = \frac{1}{r_{ij}^{1+\alpha}},
  \label{eqn:Jij}
\end{equation}
where $\alpha$ is the parameter characterizing the range of
interaction.  (Note that $\alpha$ should be positive to assure the
energy convergence.)  In spite of the extremely simple form of its
Hamiltonian, the model is known to exhibit various critical behavior
with respect to the parameter $\alpha$: When $\alpha$ is sufficiently
large, the system belongs to the same universality class as the
nearest-neighbor model, i.e., no finite-temperature phase
transitions~\cite{Ruelle1968,Dyson1969}.  At $\alpha=1$, however, the
system exhibits a Kosterlitz-Thouless phase transition at a finite
temperature~\cite{AndersonY1971,KosterlitzT1973,Kosterlitz1974}.  In
the regime $1/2 < \alpha < 1$, the critical exponents of the system
changes continuously as $\alpha$ is decreased~\cite{FisherMN1972}.
Finally, when $\alpha$ is equal to or smaller than $1/2$, the system
shows the critical exponents of the mean-field
universality~\cite{FisherMN1972}.  Such rich and nontrivial phenomena
associated with the long-range interactions have attracted much
interest and many researches have been done both theoretically and
numerically.

On numerical researches of long-range interacting spin models,
however, the standard Monte Carlo techniques encounter a serious
problem, i.e., quadratic increase of CPU time per Monte Carlo sweep as
the system size increases.  This is simply because there are ${}_NC_2
\approx N^2$ different pairs of spins to be considered in an $N$-spin
system.  For unfrustrated spin models, the cluster methods, such as
the Swendsen-Wang~\cite{SwendsenW1987} or Wolff~\cite{Wolff1989}
algorithms, are the methods of choice, since they almost completely
eliminate correlations between succeeding spin configurations on the
Markov chain.  Unfortunately, the cluster algorithms share the same
difficulty with the single spin flip update.  In 1995, however,
Luijten and Bl\"ote introduced a very efficient cluster
algorithm~\cite{LuijtenB1995}.  What they focused was that, on
average, only $O(N)$ among $O(N^2)$ bonds contribute to cluster
construction.  By employing a rejection-free method based on binary
search on a cumulative probabilities, they succeeded in reducing the
number of operations per Monte Carlo sweep drastically to $O(N \log
N)$.  Recently, the same strategy has been applied to the quantum
Monte Carlo method for long-range ferromagnetic Ising models in a
transverse external field~\cite{Sandvik2003}.

In the present paper, we propose a still faster cluster Monte Carlo
algorithm for long-range interacting ferromagnets.  Our method is
based on the extended Fortuin-Kasteleyn representation of partition
function~\cite {FortuinK1972,KawashimaG1995} and an extremely
effective technique for integral random number generation, so-called
{\em Walker's method of alias}~\cite{Walker1977,Knuth1997a}.  As the
method of Luijten and Bl\"ote, the proposed algorithm does not
introduce any cutoff for interaction range and realizes the identical
stochastic dynamics with the original $O(N^2)$ Swendsen-Wang method.
The CPU time per Monte Carlo sweep is, on the other hand, merely
proportional to $N$ instead of $N \log N$ or $N^2$, and is an order of
magnitude shorter than that of Luijten and Bl\"ote for sufficiently
large systems.  In addition to its speed, our algorithm has several
advantages: First, it is quite robust, that is, it works efficiently
both for short-range and long-range interacting models as it stands.
Second, the calculation of the total energy and the specific heat are
also possible in $O(N)$ time without any extra cost.  Third, our Monte
Carlo algorithm is straightforwardly extended for quantum models, such
as the transverse-field Ising model, the Heisenberg model, etc.

The organization of the present paper is as follows: In Sec.~2, we
briefly review the Swendsen-Wang cluster algorithm and its $O(N \log
N)$ variant by Luijten and Bl\"ote.  In Sec.~3, we present our new
algorithm in detail.  We also show how the total energy and the
specific heat are calculated in $O(N)$ time, and the extension of the
$O(N)$ algorithm to the transverse-field Ising model.  In
Sec.~4, a benchmark test of our new algorithm is presented.  As an
application of the $O(N)$ algorithm, the Kosterlitz-Thouless
transition of the Ising chain with inverse-square interaction is
investigated in Sec.~5.  Especially, we confirm the universality
between the classical and quantum Ising models in a high accuracy.
Section~6 includes a summary and discussion, followed by appendices on
some technical details about Walker's method of alias.

\section{Conventional Cluster Algorithms for Ising model with
  Long-range Interactions}

\subsection{Swendsen-Wang Method}

In this section, first we briefly review the cluster Monte Carlo method by
Swendsen and Wang~\cite{SwendsenW1987}.  Each Monte Carlo sweep of the
Swendsen-Wang algorithm consists of two procedures, graph assignment
and cluster flip.  In the former procedure, one inspects all the bonds
sequentially, and each bond is activated or deactivated with
probability
\begin{equation}
  P_{ij} =\delta_{\sigma_i^z, \sigma_j^z} [ 1-\exp(-2 \beta J_{ij}) ]
  \label{eqn:sw-pij}
\end{equation}
and $(1 - P_{ij})$, respectively.  Then, after the trials for all the
bonds, each cluster of spins connected by active bonds is flipped at
once with probability $1/2$, and a terminate configuration is
generated.

The stochastic process achieved by the Swendsen-Wang algorithm is
ergodic.  It is also proved, with the help of the Fortuin-Kasteleyn
representation of the partition
function~\cite{FortuinK1972,KawashimaG1995}, that the algorithm
satisfies the detailed balance.  The partition function of
the Hamiltonian~(\ref{eqn:hamiltonian}) is written as
\begin{equation}
  Z = \sum_c \prod_{i<j} \ e^{\beta J_{ij} \sigma_i^z \sigma_j^z} = \sum_c \prod_{\ell=1}^{N_{\rm b}} \ e^{\beta J_\ell \sigma_\ell},
  \label{eqn:pf}
\end{equation}
where $N_{\rm b}$ ($={}_NC_2$) denotes the total number of bonds,
$\ell$ is the bond index, and $J_\ell = J_{ij}$ and $\sigma_\ell =
\sigma_i^z \sigma_j^z$ are the coupling constant and the product of the
spin states of the both ends of the bond $\ell$, respectively.  We
first extend the original phase space of Ising spins $\{ c \} \ (= \{
(\sigma_1^z,\sigma_2^z,\cdots,\sigma_N^z) \})$ to the direct product of
phase spaces of spins $\{ c \}$ and {\it graphs} $\{ g \}$.  A graph
$g$ is defined by a set of variables $g_\ell$ ($\ell = 1,2, \cdots,
N_{\rm b}$), each of which is defined on each bond (or link).  The
graph variable $g_\ell$ describes whether the $\ell$th bond is
activated ($g_\ell = 1$) or not ($g_\ell = 0$).  By using the extended
phase space, the partition function~(\ref{eqn:pf}) is expressed as
\begin{equation}
  Z = C \sum_c \sum_g \omega(c, g)
  \label{eqn:pf-fk}
\end{equation}
with
\begin{equation}
  \omega(c, g) = \prod_{\ell=1}^{N_{\rm b}} \ \Delta (\sigma_\ell, g_\ell) \ V_\ell (g_\ell),
\end{equation}
where $C$ is a constant and the summation $\sum_g$ runs over
$2^{N_{\rm b}}$ possible graph configurations.  The weight functions
$\Delta$ and $V_\ell$ are defined as
\begin{align} \Delta(\sigma_\ell, g_\ell) &= \left\{
\begin{array}{l}
0 \qquad \mbox{if $g_\ell=1$ and $\sigma_\ell=-1$} \\
1 \qquad \mbox{otherwise}
\end{array}
\right. \\
V_\ell(g_\ell) &= (e^{2\beta J_\ell}-1)^{g_\ell},
\end{align}
respectively.  The equality between Eqs.~(\ref{eqn:pf}) and
(\ref{eqn:pf-fk}) is verified by figuring out the summation with
respect to $g$ in the latter:
\begin{equation}
\begin{split}
&\sum_g \omega(c,g) = \sum_g \prod_{\ell=1}^{N_{\rm b}} \Delta(\sigma_\ell, g_\ell) V_\ell(g_\ell) = \prod_{\ell=1}^{N_{\rm b}} ( \Delta (\sigma_\ell, 0) V_\ell(0) + \Delta (\sigma_\ell, 1) V_\ell(1)) \\
& \qquad = \prod_{\ell=1}^{N_{\rm b}} (1+ \delta_{\sigma_\ell, 1}(e^{2\beta J_\ell} -1) ) \\
& \qquad = e^{\beta \sum_{\ell=1}^{N_{\rm b}} J_\ell} \prod_{\ell=1}^{N_{\rm b}} (e^{-\beta J_\ell} + \delta_{\sigma_\ell, 1}(e^{\beta J_\ell} - e^{-\beta J_\ell}) ) 
= e^{\beta \sum_{\ell=1}^{N_{\rm b}} J_\ell} \prod_{\ell=1}^{N_{\rm b}} e^{\beta J_\ell \sigma_\ell},
\end{split}
\end{equation}
and thus $C=\exp(-\beta \sum_\ell J_\ell)$.  From
Eq.~(\ref{eqn:pf-fk}), we can consider $\omega(c,g)$ as a weight of
the configuration $(c, g)$ in the extended phase space.

The Swendsen-Wang method is a procedure to update the spin
configuration dynamically by going through an intermediate graph
configuration.  Consider that the initial spin configuration is $c$.
We assign a graph $g$ for the spin configuration $c$ with the
following probability:
\begin{equation}
  P(g|c) = \frac{\omega(c, g)}{\displaystyle \sum_{g'} \omega(c, g')} = \prod_{\ell=1}^{N_{\rm b}} \frac{V_\ell(g_\ell) \Delta (\sigma_\ell, g_\ell)}{\displaystyle \sum_{g'_\ell} V_\ell(g'_\ell) \Delta (\sigma_\ell, g'_\ell)}.
\end{equation}
That is, for each bond we assign $g_\ell=1$ with probability
\begin{equation}
P(g_\ell=1| \sigma_\ell) = \frac{V_\ell(1) \Delta (\sigma_\ell, 1)}{V_\ell(1) \Delta (\sigma_\ell, 1)+V_\ell(0) \Delta (\sigma_\ell, 0)} 
  = \delta_{\sigma_\ell, 1} (1- e^{-2 \beta J_\ell}).
\end{equation}
This probability turns out to be the same as the one in
Eq.~(\ref{eqn:sw-pij}).  The cluster construction in the Swendsen-Wang
algorithm is thus equivalent to assigning graph variables in the
Fortuin-Kasteleyn language.  The second procedure (cluster flip) in
the Swendsen-Wang algorithm is also represented clearly in the
Fortuin-Kasteleyn representation: Under a given graph configuration
$g$, a new spin configuration $c'$ is selected according to
probability
\begin{equation}
P(c'|g) = \frac{\omega(c', g)}{\displaystyle \sum_c \omega(c, g)} = \frac{\displaystyle \prod_{\ell=1}^{N_{\rm b}} \Delta(\sigma'_\ell, g_\ell)}{\displaystyle \sum_c \prod_{\ell=1}^{N_{\rm b}} \Delta ( \sigma_\ell, g_\ell)}.
\end{equation}
Note that $\prod_{\ell=1}^{N_{\rm b}} \Delta(\sigma_\ell, g_\ell)$
takes either 1 or 0, depending on whether all the active bonds have
$\sigma_\ell = 1$, or not.  The last equation means that among all the
allowed spin configurations, which have nonzero $\omega(c,g)$, for a
given $g$, a configuration is chosen with equal probability.  This is
equivalent to flipping each cluster of spins connected by active bonds
independently with probability $1/2$.

The detailed balance condition of the Swendsen-Wang method is thus
represented in a concrete form:
\begin{equation}
  P(c'|c) \, \omega(c) = \sum_{g} P(c'|g) P(g|c) \, \omega (c)= \sum_{g} \frac{\omega(c, g) \ \! \omega(c', g)}{\omega (g)}.
\label{eqn:sw-dbc}
\end{equation}
Since the most right-hand side of Eq.~(\ref{eqn:sw-dbc}) is symmetric
under the exchange of initial and terminal spin states $c$ and $c'$,
the detailed balance is satisfied automatically.

\subsection{$O(N \log N)$ Method by Luijten and Bl\"ote}

It has turned out that the Swendsen-Wang cluster algorithm works quite
well for wide variety of systems without frustration.  Especially, it
removes almost completely the so-called critical slowing down near the
continuous phase transition point.  Since there is no constraint about
the range of interactions in its construction, the Swendsen-Wang
algorithm is also applicable to long-range interacting systems without
any modification.  However, since the number of bonds $N_{\rm b}$ is
${}_NC_2 = O(N^2)$ in such systems, the number of operations required
for one Monte Carlo sweep is proportional to $N^2$, which is
significantly more expensive than those for the nearest-neighbor
models.

A nifty solution for reducing drastically the number of operations
from $O(N^2)$ to $O(N \log N)$ was devised by Luijten and
Bl\"ote~\cite{LuijtenB1995}.
What they noticed is separating the activation probability $P_\ell$
into the two parts:
\begin{align}
P_\ell &= p_\ell \ \delta_{\sigma_\ell, 1} \\
p_\ell &= 1- \exp( -2 \beta J_\ell).
\label{eqn:pell}
\end{align}
If one chooses candidate bonds with probability $p_\ell$ and then
activate them with probability $\delta_{\sigma_\ell, 1}$ afterward,
the probability $P_\ell$ is realized eventually.  For choosing the
candidates bonds, one could use a more efficient method than the
exhaustive search, since $p_\ell$ is independent of the spin state
$\sigma_\ell$ and predetermined statically at the beginning of the
Monte Carlo simulation.  Indeed, it is seen that the number of
candidate bonds are typically much smaller than $N_{\rm b}$.  The
average number of candidate bonds is evaluated as
\begin{equation}
  \begin{split}
  \sum_{\ell=1}^{N_{\rm b}} p_\ell &= \frac{1}{2} \sum_{i=1}^N \sum_{j \ne i} (1- e^{-2\beta J_{ij}}) \sim \frac{1}{2} \sum_{i=1}^N \int_1^{N^{1/d}} dr \ r^{d-1} (1-e^{-2\beta J(r)}) \\
&\sim \beta N \int_1^{N^{1/2}} dr \ r^{d-1} J(r).
\end{split}
\end{equation}
Here we assume the translational invariance and that $J_{ij}$ depends
only on the distance, i.e., $J_{ij}=J(r_{ij})$.  If $J(r)$ decays
faster than $r^{-d}$, which is equivalent to the condition of energy
convergence for the ferromagnetic models, the integral in the last
expression converges to a finite value in the thermodynamic limit.
Thus, at a fixed temperature, the number of candidate bonds increases
as $N$ instead of $N^2$.

For choosing candidate bonds, Luijten and Bl\"ote adopted a kind of
rejection-free method, which is based on the binary search of
cumulative probability tables.  Let us define
\begin{align}
q_m^{(0)} &= \ p_m \prod_{\ell=1}^{m-1} (1-p_\ell) \qquad (q_1^{(0)} = p_1) \\
C_m^{(0)} &= \sum_{\ell=1}^{m} q_\ell^{(0)} \qquad \mbox{($C_0^{(0)} = 0$ and $C_{N_{\rm b}+1}^{(0)} = 1$)}
\end{align}
where $q_m^{(0)}$ is the probability that the $m$th bond is eventually
chosen as a candidate after the failure for the first, second,
$\cdots$, and $(m-1)$th bonds, and $C_m^{(0)}$ is the cumulative
probability of $q_m^{(0)}$.
When an uniform real random variable $U$ ($ \in [0, 1)$) is generated,
$U$ satisfies $C_{m-1}^{(0)} \le U < C_m^{(0)}$ with probability
$q_m^{(0)}$.  The first candidate bond $m$ can then be directly chosen
by searching the first element larger than $U$.  After the $m$th bond
is activated or deactivated depending on its spin state $\sigma_m$,
one can continue the same procedure using the tables
\begin{align}
q_n^{(m)} &= \ p_n \prod_{\ell=m+1}^{n-1} (1-p_{\ell}) \qquad (q_{m+1}^{(m)} = p_{m+1}) \\
C_n^{(m)} &= \sum_{\ell=m+1}^n q_\ell^{(m)} \qquad \mbox{($C_m^{(m)} = 0$ and $C_{N_{\rm b}+1}^{(m)} = 1$).}
\end{align}
In practice, one does not have to prepare $C_n^{(m)}$ for all $m$'s,
since $C_n^{(m)}$ is readily expressed in terms of $C_m^{(0)}$ and
$q_m^{(0)}$ as
\begin{equation}
C_n^{(m)} = \frac{p_m}{q_m^{(0)}}(C_n^{(0)}-C_m^{(0)}).
\label{eqn:cnm_by_cm0}
\end{equation}
In other words, comparing $C_n^{(m)}$ to a random number $U$ ($U \in
[0,1)$) is equivalent to comparing $C_n^{(0)}$ to $C_m^{(0)} +
(q_m^{(0)}/p_m) U$.

Besides an initial table setup, which requires $O(N_{\rm b})$
operations, searching an element in the table can be performed very
quickly by using the binary search algorithm.  The number of
operations required for each search is $O(\log N_{\rm b}) = O(\log
N)$, which is significantly smaller than $O(N_{\rm b})$ for the naive
sequential search.  Since the average number of candidate bonds is
$O(N)$, a whole Monte Carlo sweep is accomplished by $O(N \log N)$
operations on average.

%%%%%%%%%%%%%%%%%%%%%%%%%%%%%%%%%%%%%%%%%%%%%%%%%%%%%%%%%%%%%%%%%%%%%%%%%%%%%%%

\section{New $\mathbf O(N)$ Cluster Algorithm}

\subsection{Formulation of $O(N)$ Method}

The factor $\log N$ in the method of Luijten and Bl\"ote is due to the
fact that they use the binary search algorithm in looking for a next
candidate.  This factor might be removed if one can use some $O(1)$
method instead of the binary search.  Walker's method of
alias~\cite{Walker1977,Knuth1997a} has been known as such an $O(1)$
method to generate integral random numbers according to arbitrary
probability distribution for a long time~(Appendix~A), and is a
potential candidate for the replacement.  Unfortunately, for the
Walker method one can not use the smart trick presented in
Eq.~(\ref{eqn:cnm_by_cm0}) for reducing the number of tables.  It
means that one has to prepare a table of length $O(N_{\rm b})$ for
each $m$ before starting the simulation.  The total amount of memory
storage for storing all the tables is thus $O(N_{\rm b}^2) = O(N^4)$,
which is not acceptable in practice.  In the following, we present a
different approach based on the extended Fortuin-Kasteleyn
representation, which solves the storage problem and enables us to use
the efficient $O(1)$ method by Walker with reasonable storage
requirement, $O(N_{\rm b})$ (or $O(N)$ for systems with translational
invariance).

Our central idea is assigning a nonnegative integer to each bond
instead of a binary (active or inactive).  The integer to be assigned
is generated according to the Poisson distribution.  The probability
that a Poisson variable takes an integer $k$ is given by
\begin{equation}
  f(k;\lambda) = \frac{e^{-\lambda} \lambda^k}{k!},
  \label{eqn:poisson}
\end{equation}
where $\lambda$ is the mean of the distribution.
Note that $f(0;\lambda)=e^{-\lambda}$ and therefore
\begin{equation}
\sum_{k=1}^\infty f(k;\lambda) = 1-e^{-\lambda},
\end{equation}
which is equal to $p_\ell$ in Eq.~(\ref{eqn:pell}), if one puts
$\lambda$ to be $2 \beta J_\ell$.  That is, if one generates an
integer according to the Poisson distribution with $\lambda = 2\beta
J_\ell$, it will take a nonzero value with probability $p_\ell$.  Thus
conventional procedure in activating bonds in the Swendsen-Wang
algorithm can be modified as follows: Generate a Poisson variable for
each bond with a mean $2\beta J_\ell$, then activate the bond only
when the variable is nonzero and the spins are parallel.
At first glance, it seems that the situation is getting worse, since a
Poisson random number, instead of a binary, is needed for each bond.
At this point, however, we leverage an important property of the
Poisson distribution: the Poisson process is that for random events
and there is no statistical correlation between each two events.  It
allows us to realize the whole distribution by calculating just one
Poisson random variable with the the mean $\lambda_{\rm tot} =
\sum_\ell \lambda_\ell$.  The following identity clearly represents
the essence:
\begin{equation}
\prod_{\ell=1}^{N_{\rm b}} f(k_\ell;\lambda_\ell) = f \left( k_{\rm tot}; \lambda_{\rm tot} \right) \frac{(k_{\rm tot} )!}{k_1! k_2! \cdots k_{N_{\rm b}}!} \prod_{\ell=1}^{N_{\rm b}} \left( \frac{\lambda_\ell}{\lambda_{\rm tot}} \right)^{k_\ell},
\label{eqn:poisson_identity}
\end{equation}
where $k_{\rm tot} = \sum_{\ell} k_\ell$.  This identity is verified
in a straightforward way by substituting Eq.~(\ref{eqn:poisson}) in
both hands.  The left-hand side of Eq.~(\ref{eqn:poisson_identity}) is
the probability that $k_\ell$ is assigned to each bond.  The
right-hand side, on the other hand, stands for the probability of
generating a single Poisson number $k_{\rm tot}$ and then distributing
$k_\ell$ events to each bond with the weight proportional to
$\lambda_\ell$.  Distributing each event can be carried out in a
constant time using Walker's method of alias.  Since generating a
Poisson number with the mean $\lambda_{\rm tot}$ takes only
$O(\lambda_{\rm tot})$ time on average, the number of operations of
the whole procedure is also proportional to $\lambda_{\rm tot} = 2
\beta \sum_\ell J_\ell$, which is $O(N)$ for energy converging models.

Before closing this section, let us describe our $O(N)$ algorithm in
terms of an extended Fortuin-Kasteleyn representation.  Introducing a
configuration $k = (k_1, k_2, \cdots, k_{N_{\rm b}})$ instead of $g =
(g_1, g_2, \cdots, g_{N_{\rm b}})$ in the original representation, the
partition function is expressed as
\begin{equation}
Z = \sum_c \sum_k \prod_{\ell=1}^{N_{\rm b}} \Delta (\sigma_\ell, k_\ell) V_\ell (k_\ell) = \sum_c \prod_{\ell=1}^{N_{\rm b}} \sum_{k_\ell=0}^{\infty} \Delta (\sigma_\ell, k_\ell) V_\ell (k_\ell)
\label{eqn:pf-efk}
\end{equation}
with
\begin{align}
\Delta(\sigma, k) &= \left\{ 
\begin{array}{l}
0 \qquad \mbox{if $k\ge 1$ and $\sigma=-1$} \\
1 \qquad \mbox{otherwise}
\end{array}
\right. \\
V_\ell(k) &= \frac{e^{- \beta J_\ell } (2\beta J_\ell)^{k}}{k!}.
\end{align}
The original partition function is easily recovered by performing the
summation over $k_\ell$'s first.

\subsection{Procedure in Concrete}
One Monte Carlo sweep of the $O(N)$ algorithm is described as follows.
\begin{enumerate}
\item Generate a nonnegative integer $k$ according to the Poisson
  distribution with the mean $\lambda_{\rm tot}$.
\item Repeat the following procedure $k$ times:
  \begin{enumerate}
    \item[(2-a)] Choose a bond $\ell$ with probability
      \begin{equation}
        \frac{J_\ell}{\sum_{\ell'=1}^{N_{\rm b}} J_{\ell'}}
      \end{equation}
      by using Walker's method of alias.
    \item[(2-b)] If $\sigma_\ell = 1$ then activate bond $\ell$.  If the bond
      is already activated, just do nothing.
    \end{enumerate}
\item Flip each cluster with probability 1/2.
\end{enumerate}
For a system with translational invariance, step~(2-a) in the above
procedure can be replaced by
\begin{quote}
\begin{enumerate}
\item[(2-a')] Choose a site $i$ with probability $1/N$, then choose
  another site $j$ with probability
   \begin{equation}
     \frac{J_{ij}}{\sum_{j'\ne i} J_{ij'}}.
   \end{equation}
\end{enumerate}
\end{quote}
In this way, the size of tables for the modified probability and alias
number in the Walker method (see Appendices A and B for details) can
be reduced from $O(N^2)$ down to $O(N)$.

\subsection{Total Energy and Specific Heat Measurement}

Measuring the total energy is also costly for long-range interacting
models.  In Ref.~\cite{KrechL2000}, Krech and Luijten proposed a
method based on the fast Fourier transform.  In this section, however,
we show that the total energy and the specific heat are also
calculated in $O(N)$ time in the present algorithm.  Indeed, the both
quantities are obtained free of charge during Monte Carlo sweeps.

Let us consider the expression for the energy in the extended
Fortuin-Kasteleyn representation.  Differentiating the partition
function~(\ref{eqn:pf-efk}) with respect to the inverse temperature,
we obtain
\begin{equation}
  \begin{split}
  E &= - \frac{\partial}{\partial \beta} \ln \Big[ \sum_c \sum_k W (c, k) \Big] 
  = \frac{\displaystyle \sum_c \sum_k \sum_\ell ( J_\ell - k_\ell/\beta ) W (c, k)}{\displaystyle \sum_c \sum_k W (c, k)} \\ & = J_{\rm tot} - \frac{1}{\beta} \Big\langle \sum_\ell k_\ell \Big\rangle_{\rm MC},
\end{split}
\label{eqn:energy}
\end{equation}
where $J_{\rm tot} = \sum_\ell J_\ell$, and $\langle \cdots
\rangle_{\rm MC}$ denotes the Monte Carlo average of an observable in
the present $O(N)$ algorithm.
Thus, in order to calculate the total energy, nothing more than the
information one uses during Monte Carlo sweeps is needed.  It also
applies to the the specific heat.  Differentiating the right-hand side
of Eq.~(\ref{eqn:energy}) once again, one obtains the following expression
\begin{equation}
  \begin{split}
  C &= - \frac{\beta^2}{N} \frac{d E}{d \beta} \\
  &= -\frac{\beta^2}{N} \left[ \frac{1}{\beta^2} \Big\langle \sum_\ell k_\ell \Big\rangle_{\rm MC} -
    \Big\langle \Big( J_{\rm tot} - \frac{1}{\beta} \sum_\ell k_\ell \Big)^2 \Big\rangle_{\rm MC} + \Big\langle J_{\rm tot} - \frac{1}{\beta} \sum_\ell k_\ell \Big\rangle^2 \right] \\
  &= \frac{1}{N} \left[ \Big\langle \Big( \sum_\ell k_\ell \Big)^2 \Big\rangle_{\rm MC} - \Big\langle \sum_\ell k_\ell \Big\rangle_{\rm MC}^2 - \Big\langle \sum_\ell k_\ell \Big\rangle_{\rm MC} \right]
  \end{split}
  \label{eqn:sheat}
\end{equation}
for the specific heat, which is not simply a variance of of
energy~(\ref{eqn:energy}) but has an extra term $\langle \sum_\ell
k_\ell \rangle_{\rm MC}$.
We note that the expressions for the total energy~(\ref{eqn:energy})
and the specific heat~(\ref{eqn:sheat}) have a close relation with
those for the quantum Monte Carlo method in the continuous
imaginary-time path integral or the high-temperature series
representations~\cite{Evertz2003}.

\subsection{Quantum Cluster Algorithm for transverse-field Ising Model}

The $O(N)$ Monte Carlo algorithm can be extended quite naturally to
quantum spin systems with long-range interactions.  In this section,
as a simplest example, we present a quantum cluster algorithm for the
long-range Ising model in a transverse external field.  Application to
other quantum spin models, such as the Heisenberg or the $XY$ models, is
also straightforward.

The Hamiltonian of the transverse-field Ising model with long-range
interactions is defined as
\begin{equation}
{\mathcal H} = - \sum_{i < j} J_{ij} \sigma_i^z \sigma_j^z - \sum_{i=1}^{N} \Gamma \sigma_i^x,
\label{eqn:tim}
\end{equation}
where $\Gamma$ denotes the strength of transverse external field, and
$\sigma_i^x$ and $\sigma_i^z$ are the Pauli operators at site $i$.
According to the standard prescription~\cite{Suzuki1976}, we start
with dividing the Hamiltonian~(\ref{eqn:tim}) into two parts, bond
terms $\mathcal{H}_{\rm b} = - \sum_{i < j} J_{ij} \sigma_i^z
\sigma_j^z$ and site terms $\mathcal{H}_{\rm s} = - \sum_{i=1}^{N}
\Gamma \sigma_i^x$.  The partition function is then expanded as
\begin{equation}
\begin{split}
Z &= {\rm Tr} \, e^{-\beta \mathcal{H}} = \lim_{M \rightarrow \infty} \sum_{\phi_1} \langle \phi_1 | \left( e^{ - \frac{\beta}{M} \mathcal{H}_{\rm b} } \ e^{ - \frac{\beta}{M} \mathcal{H}_{\rm s} } \right)^M  | \phi_1\rangle \\
&= \lim_{M \rightarrow \infty} \sum_{\phi_1, \cdots, \phi_M} \prod_{m=1}^M \ e^{-\frac{\beta}{M} E_m} \langle\phi_m |e^{-\frac{\beta}{M}\mathcal{H}_{\rm s}}|\phi_{m+1} \rangle,
\end{split}
\label{eqn:suzuki-trotter}
\end{equation}
where $M$ is the number of Suzuki-Trotter slices along the
imaginary-time axis.  In Eq.~(\ref{eqn:suzuki-trotter}), we inserted
the identities $\sum_{\phi_m} | \phi_m \rangle \langle \phi_m |$
between the operators.  The basis set $\{ \phi_m \}$ is chosen so that
$\{ \sigma^z_i \}$ are diagonalized (and so is $\mathcal{H}_{\rm b}$),
and $E_m \equiv \langle \phi_m | \mathcal{H}_{\rm b} | \phi_m
\rangle$.  We impose the periodic boundary conditions in the
imaginary-time direction: $\phi_{M+1} = \phi_1$.  Expanding the
exponential operators of the site Hamiltonian to the first order, we
obtain the following discrete imaginary-time path integral:
\begin{equation}
  \begin{split}
  Z &= \lim_{M \rightarrow \infty} \sum_{\phi_1, \cdots, \phi_M} \prod_{m=1}^M \ e^{-\frac{\beta}{M} E_m} \langle\phi_m |\prod_{i=1}^{N} \left[ 1+\frac{\beta \Gamma}{M}(\sigma_i^+ + \sigma_i^-) \right] |\phi_{m+1} \rangle \\
  &= C \lim_{M \rightarrow \infty} \sum_{\phi_1, \cdots, \phi_M} \prod_{m=1}^M \Big[ \prod_{i < j} e^{\frac{\beta J_{ij}}{M} \sigma_{i}^{(m)} \sigma_{j}^{(m)}} \Big] \Big[ \prod_{i=1}^{N} e^{\frac{1}{2} \ln \frac{\beta \Gamma}{M} \sigma_i^{(m)} \sigma_i^{(m+1)}} \Big] \\
  &= C \lim_{M \rightarrow \infty} \sum_{\phi_1, \cdots, \phi_M} \!\! \exp \Big[ \sum_{m=1}^M \sum_{i < j} \frac{\beta J_{ij}}{M} \sigma_{i}^{(m)} \sigma_{j}^{(m)} \\
  & \qquad \qquad \qquad \qquad \qquad \qquad + \sum_{m=1}^M \sum_{i=1}^{N} \frac{1}{2} \ln \frac{\beta \Gamma}{M} \sigma_i^{(m)} \sigma_i^{(m+1)} \Big],
  \end{split}
\label{eqn:discrete}
\end{equation}
where $\sigma^\pm_i = ( \sigma^x_i \pm i \sigma^y_i)/ 2$ are the spin
ladder operators, $\sigma_i^{(m)} \equiv \langle \phi_m | \sigma_i^z |
\phi_m \rangle$, and $C$ is a constant.
Thus, the partition function of the transverse-field Ising chain of
$N$ sites is represented by that of a two-dimensional classical Ising
model of $M \times N$ sites, where the interactions are long-ranged
along one axis (real space direction) and short-ranged along the other
axis (imaginary-time direction).  The coupling constants in both
directions are $\beta_{\rm cl} J^{\rm (space)}_{ij} = \beta J_{ij}/M$
and $\beta_{\rm cl} J^{\rm (time)} = \frac{1}{2} \ln (\beta
\Gamma/M)$, respectively, where $\beta_{\rm cl}$ is a fictitious
inverse temperature of the mapped system.  The $O(N)$ cluster
algorithm presented in the previous subsection is then applied to this
classical Ising model straightforwardly.

\begin{figure}
\begin{center}
\resizebox{\textwidth}{!}{\includegraphics{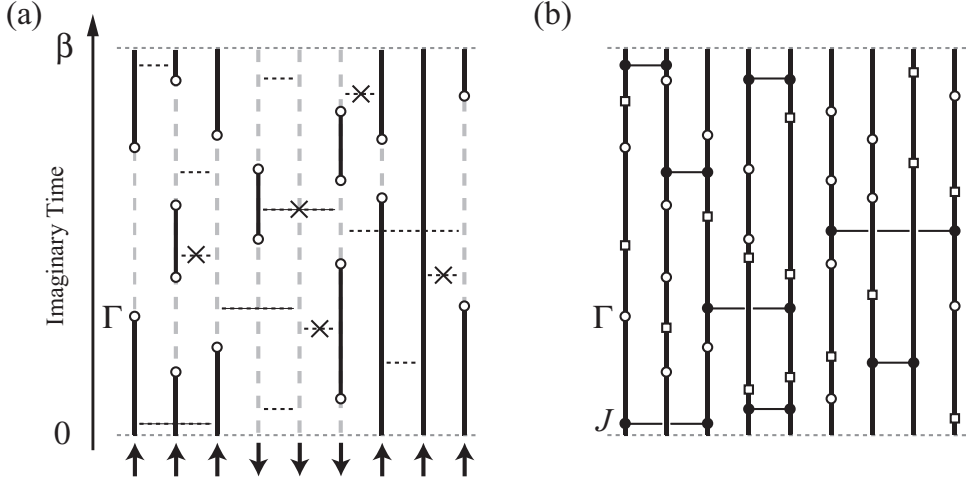}}
\vspace*{-1.0cm}
\caption{(a) Example of the space-time configuration in the
  continuous-time path integral representation.  The arrows at the
  bottom denote $\phi_0$.  Solid and broken lines denote the
  continuously aligned up and down spins, respectively.  The open
  circles represent the space-time position where an ladder operator
  is inserted.  (b) Possible graph configuration assigned to spin
  configuration~(a).  The open squares represent the positions where
  the temporal bond is deactivated, while the filled circle represent
  those where a spatial bond is activated.  The spatial
  long-range bonds are activated only when the spins are parallel at
  both ends as depicted in (a), where the candidates connecting
  antiparallel spins are rejected (x-marks).
}
\label{fig.1}
\end{center}
\end{figure}

Furthermore, it has been shown that one can take the Trotter limit ($M
\rightarrow \infty$) in Eq.~(\ref{eqn:discrete}), and perform Monte
Carlo simulations directly in the imaginary-time
continuum~\cite{BeardW1996,RiegerK1999}.  It is possible because the
coupling constant along the imaginary-time axis $J^{\rm (time)}$
increases as $M$ does.  The average number of antiparallel pairs (or
kinks) remains finite even in the continuous-time limit, and therefore
one does not have to take configurations with infinite number of kinks
into account.  Specifying the number of kinks by $n$ and its
space-time position by $(\tau_p, s_p)$ ($p = 1,2,\cdots,n$), we obtain
the continuous-time path integral representation of the partition
function:
\begin{equation}
  Z = \sum_{\phi_0} \Big[ e^{-\beta E_0}  + \sum_{n=1}^\infty \sum_{\{ s_p \}} \int_{0}^{\beta} d\tau_1 \int_{\tau_1}^\beta d\tau_2 \cdots \!\!\! \int_{\tau_{n-1}}^\beta \!\! d\tau_n \, \Gamma^n \prod_{p=1}^{n+1} e^{- (\tau_p - \tau_{p-1}) E_{p-1}} \Big],
\end{equation}
where $\tau_0 = 0$, $\tau_{n+1} = \beta$, and $E_p$ is the diagonal
energy $\langle \phi | \mathcal{H}_{\rm b} | \phi \rangle$ of the spin
configuration between the $p$th and $(p+1)$th kinks ($E_{n} = E_0$).
In Fig.~\ref{fig.1}(a) an example of path integral configuration is
shown.

The cluster algorithm is also defined directly in the Trotter limit.
Since the activation probability of temporal bond with parallel spins,
$1 - \exp (2 \beta_{\rm cl} J^{(time)}) = 1 - \beta \Gamma / M$,
becomes almost unity for $M \gg 1$, the probability of finding $n$
{\em inactive} links (open squares in Fig.~\ref{fig.1}) in a uniform
temporal segment of unit imaginary time, which contains $M/\beta$ Trotter
slices, is given by a Poisson distribution,
\begin{equation}
  _{M/\beta}C_n (1 - \beta \Gamma / M)^{(M/\beta - n)} (\beta \Gamma / M)^n \approx f(n,\Gamma).
\end{equation}
Similarly, the probability of finding $n$ spatial candidate links
(horizontal dashed lines in Fig.~\ref{fig.1}) between parallel
spins at site $i$ and $j$ in unit imaginary time is $f(n, 2J_{ij})$.
After all, the overall probability of finding $n$ {\em events} in
total at some site or bond is given by $f(n,\Lambda)$ with
\begin{equation}
  \Lambda = N\Gamma + 2 J_{\rm tot}.
\end{equation}
Since these events are statistically independent with each other, a
series of events is generated successively by using the exponential
distribution for the temporal interval $t$ between two events:
\begin{equation}
p(t) dt = \Lambda  e^{-\Lambda t} dt.
\end{equation}
At each imaginary time, then a site or bond is chosen according to the
probabilities $\Gamma / \Lambda$ or $2 J_{ij} / \Lambda$,
respectively.  This is again done in a constant time by using the
Walker method.  If a site is chosen, the temporal bond is deactivated,
i.e., clusters are disconnected at this space-time position.  If a
bond is selected (and if the spins on its ends are parallel), on the
other hand, a spatial link is inserted, i.e., two sites are connected
at this imaginary time (horizontal solid lines in Fig.~\ref{fig.1}).
At the space-time points where the spin changes its direction (open
circles in Fig.~\ref{fig.1}), we always deactivate the temporal bond.
By repeating this procedure until the imaginary time $\beta$ is
reached, the whole lattice is divided into several clusters
[Fig.~\ref{fig.1}(b)].  Finally each cluster is flipped with
probability $1/2$ to generate a terminal configuration.

The number of operations per Monte Carlo sweep is proportional to the
number of generated events.  Its average is given by $\beta \Lambda$,
which is proportional to the system size $N$ as the $O(N)$ algorithm
for classical models.  We note that the $O(N)$ quantum cluster
algorithm presented in this section is also formulated in the same way
in the high-temperature series representation~\cite{Sandvik2003}.

\section{Performance Test}

In order to demonstrate the efficiency of the present method, we
carried out Monte Carlo simulations for the classical mean-field (or
infinite-range) model of various system sizes ($N=2,4,\cdots,2^{25}$).
We use the naive Swendsen-Wang and Luijten-Bl\"ote methods as
benchmarks.  The coupling constants of the mean-field model is given
by
\begin{equation}
 J_{ij}= \frac{1}{N}
\end{equation}
for all $i \ne j$. The denominator $N$ is introduced to prevent the
energy density of the system from diverging in the thermodynamic
limit.  We choose the mean-field model as a severest test case for
these algorithms, though simpler and faster algorithms, even exact
analytic results, exist for this specific model.  The benchmark test
was performed on a PC workstation (CentOS Linux 5.1, Intel Xeon
3.2GHz, 1MB cache, GNU C++ 4.1.1).

\begin{figure}
\begin{center}
  \resizebox{!}{!}{\rotatebox{0}{\includegraphics[width=9cm]{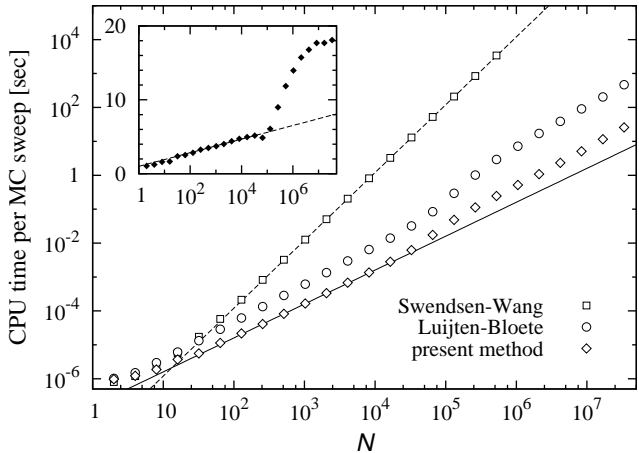}}}
  \caption{System size dependence of CPU time per Monte Carlo sweep of
    the Swendsen-Wang (squares), Luijten-Bl\"ote (circles), and the
    present (diamonds) methods for the mean-field model at $T=1$.  The
    solid and dashed lines indicate the theoretical asymptotic scaling
    for the Swendsen-Wang ($N^2$) and the present ($N$) methods,
    respectively.  The Luijten-Bl\"ote and the present methods both
    show an anomaly at $N \approx 10^5$, which is attributed to the
    occurrence of cache miss.  In the inset, the relative speed of the
    present algorithm to that of the Luijten-Bl\"ote method is also
    shown, where the dashed line indicates a $\log N$ scaling.  }
\label{fig.2}
\end{center}
\end{figure}

We confirm that all these algorithms produce the same results, total
energy, specific heat, magnetization density squared, Binder cumulant,
etc, within the error bar in the whole temperature range we simulated.
The CPU time spent for one Monte Carlo sweep at the critical
temperature ($T=1$) is shown in Fig.~\ref{fig.2}.  For the naive
Swendsen-Wang algorithm, as one expects, the CPU time grows rapidly as
$N^2$.  On the other hand, it is clearly seen in Fig.~\ref{fig.2} that
the present algorithm has a different scaling, linear to the system
size, and is indeed much faster than the Swendsen-Wang method except
for very small system sizes $(N \le 4)$.  The present and
Luijten-Bl\"ote methods exhibit a similar scaling behavior, but the
former is faster for all the system sizes we simulated.  To see the
difference in scaling behavior in detail, we plot the relative speed
of the present algorithm to the latter in the inset of
Fig.~\ref{fig.2}.  For $N \lesssim 10^5$, it scales as $\log N$, which
is consistent with the performance difference between the Walker and
the binary search algorithms.  Around the system size $N \approx
10^5$, however, the results for the present algorithm start to deviate
from the $N$-linear scaling.  Those for the Luijten-Bl\"ote method
shows a similar anomalous behavior, but the situation is much worse in
this case as seen in the inset of Fig.~\ref{fig.2}.  We attribute
these anomalies to the occurrence of cache miss, for the spin
configuration of $N \gtrsim 10^5$ does not fit the cache memory, whose
size is typically a few MB.  The naive Swendsen-Wang method should
also suffer from the same problem, but in the present benchmark test
its effect seems to be hidden under the quadratic growth in the number
of operations.

In summary, among the existing three algorithms the present $O(N)$
method is the fastest except for very small system sizes.  Especially,
it outperforms the Swendsen-Wang method by four orders of magnitude at
$N=2^{19}$ and the Luijten-Bl\"ote method by about factor twenty at
$N=2^{25}$.  This efficiency of the present method enables us to
simulate much larger systems or further improve statistics as compared
with the previous Monte Carlo studies, as demonstrated in the next
section.

\section{Kosterlitz-Thouless Transition in Ising Chain with
  Inverse-square Interactions}

In this section, as a nontrivial example, we apply our $O(N)$ cluster
algorithm to the phase transition of the one-dimensional Ising model
with inverse-square interactions [Eq.~(\ref{eqn:Jij}) with $\alpha =
1$].
As we mentioned in the introduction, among the models with
algebraically decaying interactions, this model is special as a
boundary case, i.e., it has the weakest (or shortest) interactions to
trigger a finite-temperature phase transition.  What is more, this
phase transition belongs to the same universality class as the
Kosterlitz-Thouless
transition~\cite{AndersonY1971,KosterlitzT1973,Kosterlitz1974}, where
logarithmic excitations brought by formation of domain walls compete
with the entropy generation.  The Kosterlitz-Thouless transition is
characterized by an exponential divergence of the correlation length
toward the critical temperature $T_{\rm KT}$ and a finite jump in the
magnetization.  Especially, the amount of the magnetization gap at the
critical point is conjectured to satisfy the following universal
relation:
\begin{equation}
  2 m^2 = T_{\rm KT},
  \label{eqn:universal-jump}
\end{equation}
where $m^2 = \langle ( \sum_i \sigma_i^z )^2 \rangle / N^2$, being the
square of magnetization density.  For the classical Ising chain with
inverse-square interactions, it is confirmed that a phase transition
of Kosterlitz-Thouless universality occurs by an extensive Monte Carlo
study~\cite{LuijtenM2001}.

\begin{figure}[tbp]
\begin{center}
{\small (a)} \hfill {\small (b)} \hfill \mbox{} \\[-1.5em]
\resizebox{0.49\textwidth}{!}{\includegraphics{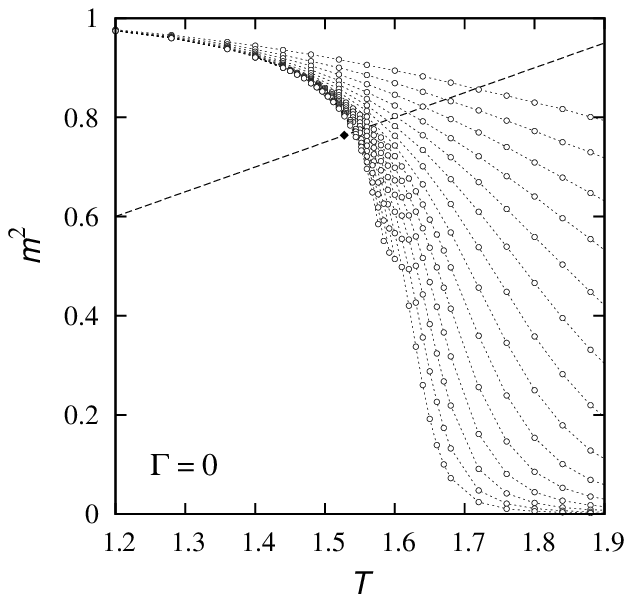}}
\hfill
\resizebox{0.49\textwidth}{!}{\includegraphics{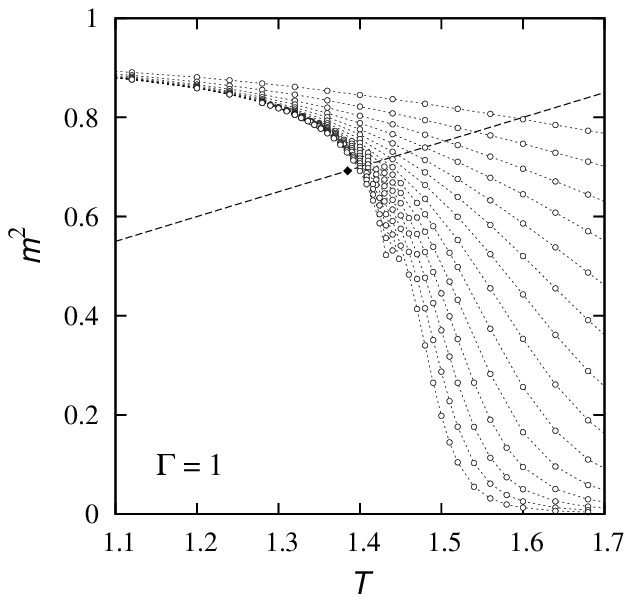}}
\caption{Temperature dependence of magnetization density squared for
  (a) classical ($\Gamma=0$) and (b) quantum ($\Gamma=1$) Ising chains
  with inverse-square interactions.  System sizes are $L=2^3$, $2^4$,
  $\cdots$, $2^{20}$ from the top to the bottom.  The error bar of
  each data point is much smaller than the symbol size.  The dashed
  lines denote the universal jump relation
  Eq.~(\ref{eqn:universal-jump}).  The filled diamond indicates the
  critical temperature obtained from the finite-size scaling analysis
  (see Fig.~\ref{fig:m2-scale} below) and the magnetization density
  squared just below the critical point.}
\label{fig:m2-raw}
\end{center}
\end{figure}

We perform Monte Carlo simulations by using the $O(N)$ cluster
algorithm for the chain length $L=2^3,2^4,\cdots,2^{20} (=1048576)$.
We impose periodic boundary conditions.  In order to minimize the
effect of boundary conditions, we use the following renormalized
coupling constant
\begin{equation}
  \tilde{J}_{ij} = \sum_{n=-\infty}^\infty \frac{1}{(i-j - nL)^2} 
  = \frac{\pi^2}{\displaystyle L^2 \sin^2 \frac{\pi (i-j)}{L}},
\end{equation}
in which contribution from all periodic images is taken into account.
It reduces to the bare coupling constant~(\ref{eqn:Jij}) in the
thermodynamic limit $L \rightarrow \infty$.  Measurement of physical
quantities is performed for 524288 Monte Carlo sweeps after discarding
8192 sweeps for thermalization.

In Fig.~\ref{fig:m2-raw}, we show the temperature dependence of the
magnetization density squared for (a) $\Gamma = 0$ and (b) $\Gamma =
1$.  For the classical system ($\Gamma = 0$), our results coincide
quite well with the previous Monte Carlo study~\cite{LuijtenM2001}.
In both cases, $m^2$ decreases monotonically as the temperature
increases.  At high temperatures, $m^2$ vanishes quite rapidly as the
system size increases, while it seems converging to a finite value in
the low temperature regime though the convergence is rather slow.
This suggests an emergence of long-range order at some finite critical
temperature.  At low temperatures, $m^2$ of the quantum system is
smaller than the classical one.  Indeed, $m^2 < 1$ even at $T=0$ for
$\Gamma=1$, which is in contrast to the classical case, $m^2 = 1$.
This is due to quantum fluctuations introduced by the transverse external
field.  In a previous quantum Monte Carlo study~\cite{Sandvik2003},
intersections of magnetization curves for different system sizes at
intermediate temperatures have been reported.  In the present study,
however, we do not observe such a nonmonotonic behavior regardless of
the system size.  We would attribute this discrepancy to a relaxation
problem in the Monte Carlo calculation in Ref.~\cite{Sandvik2003},
where only a local flip scheme is used for updating spin
configurations.

\begin{figure}[tbp]
\begin{center}
{\small (a)} \hfill {\small (b)} \hfill \mbox{} \\[-1.5em]
\resizebox{0.49\textwidth}{!}{\includegraphics{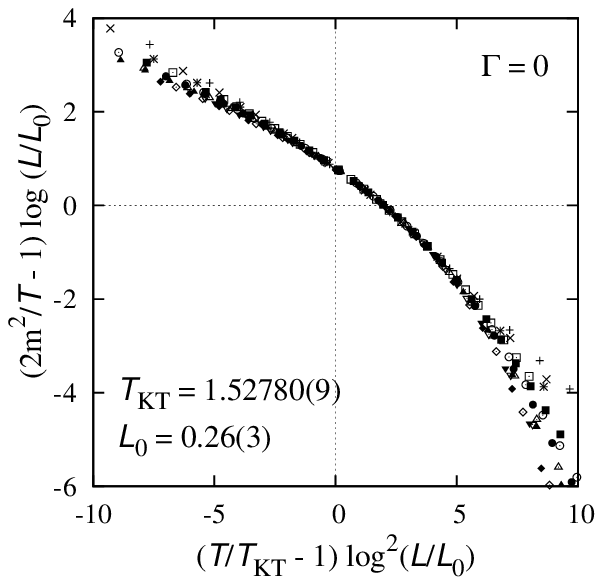}}
\hfill
\resizebox{0.49\textwidth}{!}{\includegraphics{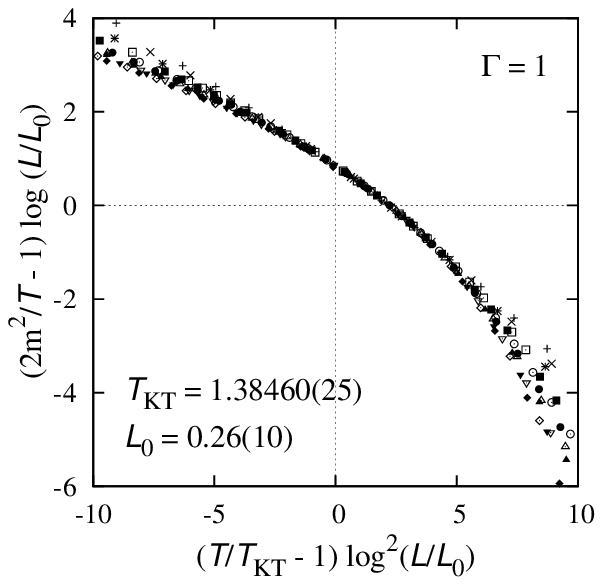}}
\caption{Scaling plot of magnetization density squared for (a)
  classical ($\Gamma=0$) and (b) quantum ($\Gamma=1$) Ising chains
  with inverse-square interactions.  System sizes are $L=2^8$
  (crosses), $2^{10}$ (x-marks), $\cdots$, $2^{20}$ (filled diamonds).
  The error bar of each data point is much smaller than the symbol
  size.}
\label{fig:m2-scale}
\end{center}
\end{figure}

In order to discuss the critical behavior in detail, next we perform a
finite-size scaling analysis.  As is well known, the standard
finite-size scaling technique does not work in the case of the
Kosterlitz-Thouless transition, for the correlation length exhibits an
exponential divergence.  Instead of the ordinary finite-size scaling,
which depends on an algebraic divergence of the correlation length, an
alternative scaling form for the magnetization has been suggested from
the renormalization group
equations~\cite{Kosterlitz1974,WeberM1987,HaradaK1997}:
\begin{equation}
  \frac{2m^2}{T} - 1 = \ell^{-1} F( t \ell^2),
  \label{eqn:fss}
\end{equation}
where $F(x)$ is a scaling function, $t = T/T_{\rm KT} - 1$, $\ell =
\log (L/L_0)$, and $L_0$ a constant.  It is confirmed that this
finite-size scaling assumption works well for the two-dimensional $XY$
model~\cite{HaradaK1997}, in which the helicity modulus, instead of
the magnetization, is the quantity exhibiting a universal jump.

In Fig.~\ref{fig:m2-scale}, we show the scaling plots for $\Gamma=0$
and 1.  Both data are scaled excellently by using the same scaling
form~(\ref{eqn:fss}), where we have only two fitting parameters,
$T_{\rm KT}$ and $L_0$.  This strongly supports that the magnetization
shows the universal jump~(\ref{eqn:universal-jump}) at the critical
point.  From these scaling plots we conclude
\begin{equation}
  T_{\rm KT} = \left\{
    \begin{array}{ll}
      1.52780(9) & \mbox{for $\Gamma = 0$} \\
      1.38460(25) & \mbox{for $\Gamma = 1$}
    \end{array}
  \right.
\end{equation}
for the Kosterlitz-Thouless critical temperature.  The result for the
classical case is compared with that in the previous Monte Carlo
study, $T_{\rm KT} = 1.5263(4)$~\cite{LuijtenM2001}, which differs
slightly beyond the error bar.  This tiny discrepancy might be due to
the difference in the way of scaling analysis.  In
Ref.~\cite{LuijtenM2001} the magnetization data at low temperatures
are first extrapolated to the thermodynamic limit, then further
extrapolated towards the critical point, whereas the Monte Carlo data
are directly used to estimate the critical temperature in the present
finite-size scaling analysis.  Thus, we expect that the present
estimate for $T_{\rm KT}$ is more reliable.

As for the quantum system ($\Gamma = 1$), the critical temperature is
lower than the classical one due to the quantum fluctuations.
However, the finite-scaling analysis confirms that the phase
transition belongs to the Kosterlitz-Thouless universality class as in
the classical case.  The finite-size scaling plots shown in
Fig.~\ref{fig:m2-scale} suggest that the scaling function itself is
universal as well.  As $\Gamma$ is increased further, $T_{\rm KT}$
decreases monotonically, and it finally vanishes at a critical
transverse field $\Gamma_{\rm c}$ ($\approx 2.52$~\cite{TodoF2100}),
where a quantum phase transition occurs.  At this point some exotic
quantum critical behavior with a nontrivial dynamical exponent $z$ is
expected, since the $(1+1)$-dimensional system is extremely
anisotropic, i.e., in real space direction the system has long-range
interactions, whereas the interaction in the imaginary-time direction
is still short ranged.  More detailed analyses on the quantum
criticality of the transverse-field Ising model will be presented
elsewhere~\cite{TodoF2100}.

\section{Summary and Discussion}

We presented an $O(N)$ cluster algorithm for Ising models with
long-range interactions.  The algorithm proposed in the present paper
is {\em exact}, i.e., it does not introduce any cutoff for interaction
range and thus it strictly fulfills the detailed balance.  Our
algorithm is formulated based on the extended Fortuin-Kasteleyn
representation, where bond variables have a nonnegative integral value
instead of a binary number.  For each bond, an integer is generated
according to the Poisson distribution.  However, it does not
necessarily mean that each Poisson variable has to be generated one by
one.  We show that generating an overall Poisson distribution and
ex-post assignment of events, using Walker's method of alias, are
statistically equivalent to the naive Swendsen-Wang method.  In
Sec.~4, we demonstrated the $N$-linear scaling behavior in the CPU
time for the mean-field model.

The present method has several advantages over the existing methods,
such as the Metropolis method, the Swendsen-Wang
algorithm~\cite{SwendsenW1987}, the improvement by Luijten and
Bl\"ote~\cite{LuijtenB1995}, or the recently proposed $O(N)$
method~\cite{SasakiM07}, in several aspects: (a)~The CPU time per
Monte Carlo sweep is $O(N)$.  (b)~It works effectively both for
short-range and long-range interacting models.  (c)~It is a cluster
algorithm and free from the critical slowing down near the critical
point.  (d)~It is possible to formulate a single-cluster
variant~\cite{Wolff1989}. (e)~It is very easy to implement the
algorithm, based on an existing Swendsen-Wang code.  (f)~It can also
be used for systems without translational invariance, though it once
costs $O(N^2)$ to initialize lookup tables.  (g)~Calculation of the
total energy and the specific heat can be done all together in $O(N)$
time.  (h)~It can be applied to Potts, $XY$, and Heisenberg models
with the help of Wolff's embedding technique~\cite{Wolff1989}.
(i)~Extension to quantum models, such as the transverse-field Ising
model or the Heisenberg model, is also possible straightforwardly.

In Sec.~5, we have applied our new algorithm to the phase transition
of Ising model with inverse square interactions, where we see that the
$O(N)$ method works ideally for both of the classical and quantum
systems.  It is confirmed in a high accuracy that the phase transition
belongs to the same universality as the Kosterlitz-Thouless
transition.

Finally, let us discuss the efficiency of the present algorithm at
very low temperatures.  For a fixed system size $N$, the calculation
cost of the present method grows linearly as the inverse temperature
$\beta$ increases, whereas that of the naive method is constant
regardless of the temperature for classical Ising models.  It
indicates that at lower temperatures than some threshold $1/\beta_{\rm
  thresh}$, the naive method outperforms the present method.  At
extremely low temperatures, almost all the bonds are activated.  The
present method then activates such bonds many times, which is the
cause of the slowing down.  Although the $\beta$-linear increase of
CPU time is inevitable for quantum systems, where the standard quantum
Monte Carlo algorithms for short-range models also suffer from the
same slowing down, however, one can adopt a ``hybrid'' scheme to
optimize the calculation cost at intermediate temperatures,
$\max(J_{\ell}) \lesssim \beta < \beta_{\rm thresh}$ for classical
models.  Suppose $J_{\ell}$'s are sorted in descending order, and we
use the naive method for the first $n$ bonds and the $O(N)$ method for
the others.  The CPU time $C(n)$ per Monte Carlo sweep is estimated as
\begin{equation}
C(n) \simeq A n + B \! \sum_{\ell=n+1}^{N_{\rm b}} \! 2 \beta J_\ell,
\end{equation}
where $A$ and $B$ are some constants.  The optimal value of $n$ is
then given by $\Delta C = C(n+1) - C(n) = A - 2 B \beta J_n = 0$.  For
the one-dimensional model with algebraically decaying
interactions~(\ref{eqn:Jij}), for example, we have
\begin{equation}
\frac{n_{\rm opt}}{N_{\rm b}} \approx N^{-1} \Big( \frac{2 B \beta}{A} \Big)^{\frac{1}{1+\alpha}}.
\end{equation}
The threshold $\beta_{\rm thresh}$ is defined as the inverse
temperature where $n_{\rm opt} = N_{\rm b} \simeq N^2$, that is,
$\beta_{\rm thresh} \approx (A/2B) N^{1+\alpha}$, which grows as the
system size $N$ increases.

\begin{ack}
  The part of the simulations in the present paper has been done by
  using the facility of the Supercomputer Center, Institute for Solid
  State Physics, University of Tokyo.  The simulation code has been
  developed based on the ALPS/looper
  library~\cite{LOOPERweb,TodoK2001,ALPS2007}.  The authors
  acknowledge support by Grant-in-Aid for Scientific Research Program
  (No.15740232) from JSPS, and also by the Grand Challenge to
  Next-Generation Integrated Nanoscience, Development and Application
  of Advanced High-Performance Supercomputer Project from MEXT, Japan.
\end{ack}

\appendix

\section{Walker's Method of Alias}

Consider a random variable $X$ which takes an integral value $i$
according to a probability $p_i$ ($1\le i \le N$ and $\sum p_i = 1$).
In this appendix, we discuss how to generate such random numbers
effectively.  One of the simplest and the most well-known methods is
the one based on rejection:

Rejection Method
\begin{enumerate}
\item Generate a uniform integral random variable $M$ ($1 \le M \le N$).
\item Generate a uniform real random variable $U$ ($0 \le U < 1$).
\item If $U$ is smaller than $p_M/p_{\mbox{max}}$ then $X=M$,
  otherwise repeat from (1).
\end{enumerate}
Here, $p_{\mbox{max}} = \max(p_i)$.  Since the acceptance rate in step
(3) is $1 / (N p_{\mbox{max}})$ ($\equiv q$), the probability of
obtaining $X=i$ eventually is $\sum_{r=1}^\infty p_i (1-q)^{r-1} q =
p_i$.  Notice that the number of iterations is $\sum_{r=1}^\infty r
(1-q)^{r-1} q = 1/q = N p_{\mbox{max}}$ on average, and therefore it
would take $O(N)$ time for each generation.  Especially, the
efficiency decreases quite rapidly as the variance of $p_i$ increases.
One may reduce the number of operations down to $O(\log N)$ by
employing the binary search on the table of cumulative probabilities
(see Sec.~2.2).  However, there exists a further effective method,
called ``Walker's method of alias''~\cite{Walker1977,Knuth1997a},
which is rejection free and generates a random integer in a constant
time.

The Walker algorithm requires two tables of size $N$, which need to be
calculated in advance.  One is the table of integral alias numbers
$\{A_i\}$ ($1 \le A_i \le N$) and the other is that of modified
probabilities $\{P_i\}$ ($0 \le P_i \le 1$).  Using these tables a
random integer is generated by the following procedure:

Walker's Method of Alias
\begin{enumerate}
\item Generate a uniform integral random variable $M$ ($1 \le M \le N$).
\item Generate a uniform real random variable $U$ ($0 \le U < 1$).
\item If $U$ is smaller than $P_M$ then $X=M$, otherwise $X=A_M$.
\end{enumerate}
This procedure has no iterations, and thus completes in a constant
time.  The meaning of the tables $\{ A_i \}$ and $\{ P_i \}$ and the
correctness of the algorithm is readily understood with the following
example:
\begin{center}
\begin{tabular}{cccccccccccccccc}
$i$   & 1 & 2               & 3            & 4            & 5            & 6            & 7     & 8    & 9    & 10   & 11   & 12   \\
$p_i$ & 0 &$\frac{1}{36}$&$\frac{2}{36}$&$\frac{3}{36}$&$\frac{4}{36}$&$\frac{5}{36}$&$\frac{6}{36}$&$\frac{5}{36}$&$\frac{4}{36}$&$\frac{3}{36}$&$\frac{2}{36}$&$\frac{1}{36}$ \\
$P_i$ &0&$\frac{1}{3}$&$\frac{2}{3}$&$\frac{3}{3}$&$\frac{3}{3}$&$\frac{2}{3}$&$\frac{2}{3}$&$\frac{1}{3}$&$\frac{2}{3}$&0&$\frac{2}{3}$&$\frac{1}{3}$ \\
$A_i$ & 10  & 9    & 8    & *    & *   & 5    &  6   &  6   &  7   & 7    & 8    & 8
\end{tabular}
\end{center}
The modified probabilities $P_i$ are determined from $p_i$ (see
Appendix~B), which gives the probabilities whether one should accept
the firstly chosen number or choose the alias number $A_i$.
Let us consider, for example, the probability of $X=9$.  There are two
possibilities: One is $M=9$ and $U < P_9$, and the other is $M=2$ and
$U \ge P_2$ since $A_2 = 9$.  The sum of these two probabilities is
\begin{equation}
  \frac{1}{12} [ P_9 + (1-P_2) ] = \frac{1}{9},
  \label{eqn:a-1}
\end{equation}
which is equal to $p_9$ as expected.  One can confirm that $\{ P_i \}$
and $\{ A_i \}$ are given in the example so that
\begin{equation}
  p_i = \frac{1}{N} \Big[ P_i + \sum_{j = 1}^N (1-P_j) \delta_{i, A_j} \Big]
  \label{eqn:a-2}
\end{equation}
holds for $i=1,2,\cdots,N$.  Together with the ordinary requirement
for probabilities, $0 \le P_i \le 1$ ($i=1,2,\cdots,N$),
Eq.~(\ref{eqn:a-2}) is the necessary condition for $\{ P_i \}$ and $\{
A_i \}$ to satisfy.  

In practice, when $N$ is not a power of two, we expand the size of
tables from $N$ to $N_{\rm opt}$, where $N_{\rm opt}$ is the smallest
integer satisfying $N_{\rm opt} \ge N$.  For $N+1 \le i \le N_{\rm
  opt}$, we assume $p_i = 0$.  In this way, generating $M$ in step~(1)
is optimized as a bit shift operation on a 32- or 64-bit integral
random number~\cite{Knuth1997a}.  Furthermore, steps~(2) and (3) can be
replaced by a comparison between two integral variables by preparing a
table of integers $\{ 2^{32} P_i\}$ (or $\{ 2^{64} P_i \}$) instead of
floating point numbers $\{ P_i \}$, by which a costly conversion from
an integer to a floating point variable can also be avoided.

In summary, by using the Walker method, integral random numbers
according to arbitrary probabilities can be generated in a constant
time.  This extreme efficiency is essential for the present $O(N)$
cluster Monte Carlo method.  In the next appendix, we describe how to
prepare the tables $\{ P_i \}$ and $\{ A_i \}$.

\section{Preparation of Modified Probabilities and Aliases}

In the original paper by Walker~\cite{Walker1977} and also in the
standard literature~\cite{Knuth1997a}, only a naive $O(N^2)$ method is
presented for initializing $\{ P_i \}$ and $\{ A_i \}$.  Here we
propose for the first time an efficient alternative procedure, which
takes only $O(N)$ time.

Consider the following table of probabilities for $\{ p_i \}$:
\begin{center}
\begin{tabular}{cccccccccccccccc}
$i$   & 1 & 2               & 3            & 4            & 5            & 6            & 7     & 8    & 9    & 10   & 11   & 12   \\
$p_i$  & 0 &$\frac{1}{36}$&$\frac{2}{36}$&$\frac{3}{36}$&$\frac{4}{36}$&$\frac{5}{36}$&$\frac{6}{36}$&$\frac{5}{36}$&$\frac{4}{36}$&$\frac{3}{36}$&$\frac{2}{36}$&$\frac{1}{36}$ \\
$P_i$ & 0 &$\frac{1}{3} $&$\frac{2}{3} $&$\frac{3}{3} $&$\frac{4}{3} $&$\frac{5}{3} $&$\frac{6}{3} $&$\frac{5}{3} $&$\frac{4}{3} $&$\frac{3}{3} $&$\frac{2}{3} $&$\frac{1}{3} $
\end{tabular}
\end{center}
Here $P_i$ is initially set to a tentative value $N p_i$ for
$i=1,\cdots,N$.  First we rearrange the table so that all the elements
with $P_i \ge 1$ precede those with $P_i < 1$
\begin{center}
\begin{tabular}{cccccccc|ccccc}
& & & & & & & $\triangledown$ & & & & & $\blacktriangledown$ \\
$i$    & 4 & 5               & 6            & 7            & 8            & 9            & 10    & 12   & 11   & 3    & 2    & 1    \\
$P_i $ & $\frac{3}{3}$ &$\frac{4}{3} $&$\frac{5}{3} $&$\frac{6}{3} $&$\frac{5}{3} $&$\frac{4}{3} $&$\frac{3}{3} $&$\frac{1}{3} $&$\frac{2}{3} $&$\frac{2}{3} $&$\frac{1}{3} $& 0
\end{tabular}
\end{center}
The rearrangement can be done by $N$ steps in contrast to the perfect
sorting, which is an $O(N \log N)$ procedure.  The white triangle
points to the rightmost element with $P_i \ge 1$ and the black
triangle points to the rightmost element in the rearranged table.

Next, we determine the alias numbers $A_i$ sequentially from the
right.  We fill the ``shortfall'' $(1-P_i)$ of the element pointed by
the solid triangle by the one pointed by the white triangle.  The
latter is always large enough, since $P_i \ge 1$ by definition.  In
the present example, first the shortfall of the rightmost element
$(1-P_1) = 1$ is filled by the element with $i=10$.  The alias number
for $i=1$ is then set to 10 and $P_{10}$ is replaced by $P_{10} -
(1-P_1) = 0$.  Since $P_{10}$ is no more larger than nor equal to
unity, we shift the white triangle to the left by one.  We repeat the
same for the next ``unfilled'' element.  After four iterations, the
table is transformed as follows:
\begin{center}
\begin{tabular}{ccccc|cccc|cccc}
& & & & $\triangledown$ & & & & $\blacktriangledown$ \\
$i$    & 4 & 5               & 6            & 7            & 8            & 9            & 10    & 12   & 11   & 3    & 2    & 1    \\
$P_i $ & $\frac{3}{3}$ &$\frac{4}{3} $&$\frac{5}{3} $&$\frac{6}{3} $&$\frac{1}{3} $&$\frac{2}{3} $&$0$&$\frac{1}{3} $&$\frac{2}{3} $&$\frac{2}{3} $&$\frac{1}{3} $& 0 \\
$A_i$ & & & & & & & & & 8 & 8 & 9 & 10
\end{tabular}
\end{center}
Here the solid and white triangles are shifted four and three times
from their original positions, respectively.  The above procedure is
repeated until the black triangle points to the same element as the
white one, i.e., all the elements get filled.  One should note that
the black triangle always moves by one after each iteration, though
the white one may stay on the same element depending whether $P_i \ge
1$ or not after the step.  The whole procedure is thus completed at
most after $(N-1)$ iterations.  In the present example, after 10
iterations we end up with
\begin{center}
\begin{tabular}{ccc|cccccccccc}
& & $\blacktriangledown$ \\[-1em]
& & $\triangledown$ \\[-.5em]
$i$    & 4 & 5               & 6            & 7            & 8            & 9            & 10    & 12   & 11   & 3    & 2    & 1    \\
$P_i $ & $\frac{3}{3}$ &$\frac{3}{3} $&$\frac{2}{3} $&$\frac{2}{3} $&$\frac{1}{3} $&$\frac{2}{3} $&$0$&$\frac{1}{3} $&$\frac{2}{3} $&$\frac{2}{3} $&$\frac{1}{3} $& 0 \\
$A_i$ & * & * & 5 & 6 & 6 & 7 & 7 & 8 & 8 & 8 & 9 & 10
\end{tabular}
\end{center}
which is equivalent to the table presented in Appendix~A.

\bibliography{main}

\begin{thebibliography}{10}
\expandafter\ifx\csname url\endcsname\relax
  \def\url#1{\texttt{#1}}\fi
\expandafter\ifx\csname urlprefix\endcsname\relax\def\urlprefix{URL }\fi

\bibitem{Ruelle1968}
D.~Ruelle, Statistical mechanics of a one-dimensional lattice gas, Commun.
  Math. Phys. 9 (1968) 267.

\bibitem{Dyson1969}
F.~J. Dyson, Existence of a phase-transition in a one-dimensional {Ising}
  ferromagnet, Commun. Math. Phys. 12 (1969) 91.

\bibitem{AndersonY1971}
P.~W. Anderson, G.~Yuval, Some numerical results on the {Kondo} problem and the
  inverse square one-dimensional {Ising} model, J. Phys. C 4 (1971) 607.

\bibitem{KosterlitzT1973}
J.~M. Kosterlitz, D.~J. Thouless, Ordering, metastability and phase transitions
  in two-dimensional systems, J. Phys. C 6 (1973) 1181.

\bibitem{Kosterlitz1974}
J.~M. Kosterlitz, The critical properties of the two-dimensional {XY} model, J.
  Phys. C 7 (1974) 1046.

\bibitem{FisherMN1972}
M.~E. Fisher, S.-K. Ma, B.~G. Nickel, Critical exponents for long-range
  interactions, Phys. Rev. Lett. 29 (1972) 917.

\bibitem{SwendsenW1987}
R.~H. Swendsen, J.~S. Wang, Nonuniversal critical dynamics in monte carlo
  simulations, Phys. Rev. Lett. 58 (1987) 86.

\bibitem{Wolff1989}
U.~Wolff, Collective monte carlo updating for spin systems, Phys. Rev. Lett. 62
  (1989) 361.

\bibitem{LuijtenB1995}
E.~Luijten, H.~W.~J. Bl\"ote, {Monte} {Carlo} method for spin models with
  long-range interactions, Int. J. Mod. Phys. C 6 (1995) 359.

\bibitem{Sandvik2003}
A.~W. Sandvik, Stochastic series expansion method for quantum {Ising} models
  with arbitrary interactions, Phys. Rev. E 68 (2003) 056701.

\bibitem{FortuinK1972}
C.~M. Fortuin, P.~W. Kasteleyn, On the random-cluster model {I.} introduction
  and relation to other models, Physica 57 (1972) 536.

\bibitem{KawashimaG1995}
N.~Kawashima, J.~E. Gubernatis, Generalization of the {Fortuin-Kasteleyn}
  transformation and its application to quantum spin simulations, J. Stat.
  Phys. 80 (1995) 169.

\bibitem{Walker1977}
A.~J. Walker, An efficient method for generating discrete random variables with
  general distributions, ACM Trans. Math. Software 3 (1977) 253.

\bibitem{Knuth1997a}
D.~E. Knuth, The Art of Computer Programming, 3rd Edition, Vol. 2,
  Seminumerical Algorithms, Addison Wesley, Reading, 1997, p. 119.

\bibitem{KrechL2000}
M.~Krech, E.~Luijten, Optimized energy calculation in lattice systems with
  long-range interactions, Phys. Rev. E 61 (2000) 2058.

\bibitem{Evertz2003}
H.~G. Evertz, The loop algorithm, Adv. in Physics 52 (2003) 1.

\bibitem{Suzuki1976}
M.~Suzuki, Relationship between $d$-dimensional quantal spin systems and
  $(d+1)$-dimensional {Ising} systems, Prog. Theor. Phys. 56 (1976) 1454.

\bibitem{BeardW1996}
B.~B. Beard, U.~J. Wiese, Simulations of discrete quantum systems in continuous
  {Euclidean} time, Phys. Rev. Lett. 77 (1996) 5130.

\bibitem{RiegerK1999}
H.~Rieger, N.~Kawashima, Application of a continuous time cluster algorithm to
  the two-dimensional random quantum {Ising} ferromagnet, Eur. Phys. J. B 9
  (1999) 233.

\bibitem{LuijtenM2001}
E.~Luijten, H.~Me{\ss}ingfeld, Criticality in one dimension with inverse
  square-law potential, Phys. Rev. Lett. 86 (2001) 5305.

\bibitem{WeberM1987}
H.~Weber, P.~Minnhagen, Monte carlo determination of the critical temperature
  for the two-dimensional {XY} mode, Phys. Rev. B 37 (1987) 5986.

\bibitem{HaradaK1997}
K.~Harada, N.~Kawashima, Universal jump in the helicity modulus of the
  two-dimensional quantum {XY} model, Phys. Rev. B 55 (1997) R11949.

\bibitem{TodoF2100}
S.~Todo, K.~Fukui, unpublished.

\bibitem{SasakiM07}
M.~Sasaki, F.~Matsubara, Stochastic cutoff method for long-range interacting
  systems   arXiv:0710.1177.

\bibitem{LOOPERweb}
{\tt http://wistaria.comp-phys.org/alps-looper/}.

\bibitem{TodoK2001}
S.~Todo, K.~Kato, Cluster algorithms for general-{S} quantum spin systems,
  Phys. Rev. Lett. 87 (2001) 047203.

\bibitem{ALPS2007}
A.~Albuquerque, F.~Alet, P.~Corboz, P.~Dayal, A.~Feiguin, S.~Fuchs, L.~Gamper,
  E.~Gull, S.~G\"urtler, A.~Honecker, R.~Igarashi, M.~K\"orner, A.~Kozhevnikov,
  A.~L\"auchli, S.~Manmana, M.~Matsumoto, I.~McCulloch, F.~Michel, R.~Noack,
  G.~Pawlowski, L.~Pollet, T.~Pruschke, U.~Schollw\"ock, S.~Todo, S.~Trebst,
  M.~Troyer, P.~Werner, S.~Wessel, The {ALPS} project release 1.3: Open-source
  software for strongly correlated systems, J. Mag. Mag. Mat. 310 (2007) 1187.

\end{thebibliography}

\end{document}